\documentclass[12pt,showpacs,preprintnumbers,superscriptaddress,amsmath,amssymb,nofootinbib]{revtex4}
\usepackage{graphicx}
\usepackage{dcolumn}
\usepackage{bm}
\usepackage{amssymb}
\usepackage{amsmath}
\usepackage{epsfig}    
\usepackage{color}
\usepackage{hhline}

\def\be{\begin{equation}}
\def\ee{\end{equation}}
\newcommand{\bea}{\begin{eqnarray}}
\newcommand{\eea}{\end{eqnarray}}
\newcommand{\nn}{\nonumber}

\numberwithin{equation}{section}

\begin{document}

{\begin{flushright}{KIAS-P180xx}
\end{flushright}}

\title{A radiative neutrino mass model in light of DAMPE excess\\ with hidden gauged  $U(1)$ symmetry }
%

\author{Takaaki Nomura}
\email{nomura@kias.re.kr}
\affiliation{School of Physics, KIAS, Seoul 02455, Republic of Korea}

\author{Hiroshi Okada}
\email{macokada3hiroshi@cts.nthu.edu.tw}
\affiliation{Physics Division, {National Center for Theoretical Sciences}, Hsinchu, Taiwan 300}

\author{Peiwen Wu}
\email{pwwu@kias.re.kr}
\affiliation{School of Physics, KIAS, Seoul 02455, Republic of Korea}

\date{\today}

\begin{abstract}
We propose a one-loop induced neutrino mass model with hidden $U(1)$ gauge symmetry, in which
we successfully involve a bosonic dark matter (DM) candidate propagating inside a loop diagram in neutrino mass generation
to explain the $e^+e^-$ excess recently reported by the DArk Matter Particle Explorer (DAMPE) experiment. 
In our scenario dark matter annihilates into four leptons through $Z'$ boson as DM DM $\to Z' Z' (Z' \to \ell^+ \ell^-)$ and $Z'$ decays into leptons via one-loop effect.
We then investigate branching ratios of $Z'$ taking into account lepton flavor violations and neutrino oscillation data.
\end{abstract}
\maketitle
\newpage

\section{Introduction}
The excess of positron/electron reported by the recent experiment of DArk Matter Particle Explorer (DAMPE)~\cite{dampe,Ambrosi:2017wek}, 
shows us several unique features if its excess could originate as a fundamental dark matter (DM) particle~\cite{Yuan:2017ysv}; 
\begin{enumerate}   
\item The excess is monochromatically sharp, as if DM directly annihilates into a pair of  electron and positron.  
\item The distribution energy is $\sim$1.4 TeV implying that the DM mass can be estimated to be $\sim$1.5 TeV in case of two body annihilation and $\sim$3 TeV in case of four body annihilation.  
\item The needed cross section is $\gtrsim 1\times10^{-26}$ cm$^3$/s ($\simeq 1\times 10^{-9}$ GeV$^{-2}$), when one assumes an existence of dark subhalo near the earth,
 \end{enumerate}
where we have focussed on an annihilating thermal DM scenario.
The third one suggests that the scale of the DM annihilation cross section is close to the one to explain the relic density of DM $\sim$0.12~\cite{Ade:2013zuv}. As a consequence, one might result in s-wave dominant DM. 
To interpret the excess, many papers have already appeared \cite{Fan:2017sor, Gu:2017gle, Duan:2017pkq, Zu:2017dzm, Tang:2017lfb, Chao:2017yjg, Gu:2017bdw, Athron:2017drj, Cao:2017ydw, Duan:2017qwj, Liu:2017rgs, Chao-guo-li-shu, Huang:2017egk, Gao:2017pym, Niu:2017hqe, Chen:2017tva, Li:2017tmd, Zhu:2017tvk, Gu:2017lir, Nomura:2017ohi, Ghorbani:2017cey, Cao:2017sju, Yang:2017cjm, Ding:2017jdr, Liu:2017obm, Ge:2017tkd, Tykhonov:2017uno, Xu:2017yix, Zhao:2017nrt, Sui:2017qra, Okada:2017pgr,Chao:2017emq,Cao:2017rjr,Han:2017ars,Niu:2017lts,Fowlie:2017fya}. We also noticed an interesting DM study through astrophysical signals \cite{Profumo:2017obk} appearing shortly before the DAMPE results.

In this letter, we propose a bosonic DM candidate that also plays a role in inducing the active neutrino masses, lepton flavor violations
(LFVs), and muon anomalous magnetic moment ($\Delta a_\mu$) at one-loop level, by introducing hidden gauged $U(1)_H$ symmetry~\cite{Ko:2016sxg, Nomura:2017wxf}.
In our scenario, the SM fields are not charged under the $U(1)_H$ at tree level while extra particles including DM candidate are charged.
The DAMPE excess as well as relic density is expected to be induced from a contact interaction via $Z'$ boson with four-body lepton annihilation and then their final state electron/positron pairs are generated at one-loop level via Yukawa coupling; therefore $2X\to 2Z'\to \bar\ell\ell\bar\ell\ell$
($\ell=e,\mu,\tau,\nu_e,\nu_\mu,\nu_\tau$). Since its numerical value of Yukawa coupling is determined by the neutrino oscillation data, LFVs, and the sizable $\Delta a_\mu$, we can hopefully predict their ratio of final state of leptons to some extent.
This is one of the big motivations to work in such a framework.

This paper is organized as follows.
In Sec.~II, we introduce our model, and then formulate each of sector such as Higgs sector, active neutrinos, LFVs, $\Delta a_\mu$ and branching ratio of $Z'$ boson, and the DM sector.
In Sec.~III, we show branching ratio of $Z'$ in allowed parameter region to satisfy neutrino oscillation data and constraints from LFVs through the numerical analysis. 
Finally we carry out numerical analysis for the DM annihilation cross section and show that the DAMPE excess as well as relic density can be explained.
We conclude in Sec.~IV.

\section{The Model}
 \begin{widetext}
\begin{center} 
\begin{table}[h]
\begin{tabular}{|c||c||c|c|c|}\hline\hline  
 & ~$L'_{a}$~ & ~$\varphi$~ & ~$\Delta$~ & ~$S$~
\\\hline 
 $SU(2)_L$ & $\bm{2}$ & $\bm{1}$ & $\bm{2}$ & $\bm{1}$   \\\hline 
$U(1)_Y$ & $-\frac12$   & $0$ & $1$  & $0$ \\\hline
 $U(1)_H $ & $x$  & $4x$   & $-x$ & $-x$    \\\hline
\end{tabular}
\caption{New field contents of fermions and bosons
and their charge assignments under $SU(2)_L\times U(1)_Y\times U(1)_H$, where the lower index $a(=1-3)$ is the number of flavors and $x$ is an arbitrary value with nonzero.}
\label{tab:1}
\end{table}
\end{center}
\end{widetext}

In this section we introduce our model and derive some formulas such as neutrino mass matrix, lepton flavor violation, $Z'$ boson interactions and DM interactions.

\subsection{Model setup}
In the fermion sector, we introduce three isospin doublet fermions $L'\equiv[N,E]^T$ with $x$ under $U(1)_H$ charge.
In the boson sector, we introduce two types inert bosons $S$ and $\Delta$ with $-x(\neq0)$ under $U(1)_H$ symmetry,
where $S$ is an isospin singlet and $\Delta$ is an isospin triplet.
Another isospin singlet $\varphi$ with $4x$ under $U(1)_H$ symmetry has nonzero vacuum expectation values (VEVs),
which is denoted by $\langle\varphi\rangle\equiv v'/\sqrt2$, where 
the VEV of  SM Higgs $H$ is denoted by $\langle H\rangle\equiv v/\sqrt2$.
All the new field contents and their assignments are summarized in table~\ref{tab:1}.
Notice here that {\it all the SM fields are not charged under $U(1)_H$ symmetry.}
The relevant Yukawa Lagrangian to generate the neutrino masses and scalar potential under these assignments
are given by
\begin{align}
&-\mathcal{L}_{Y}
 \supset
 f_{i\alpha} \bar L_{L_i} L'_{R_\alpha} S +  g_{\alpha j} \bar L'_{L_\alpha} \epsilon \Delta^*   L^c_{L_j}
  + M_\alpha \bar L'_{L_\alpha} L'_{R_\alpha} +\rm{h.c.}, \label{eq:Lag} \\
&  V \supset \mu_\varphi^2 |\varphi|^2 + m_\Delta^2 {\rm Tr}[\Delta^\dagger \Delta] + m_S^2 |S|^2 + \lambda_{\varphi} |\varphi|^4 + \lambda_{S} |S|^4 
+ \lambda_\Delta {\rm Tr}[\Delta^\dagger \Delta ]^2 + \lambda'_\Delta {\rm Tr}[(\Delta^\dagger \Delta)^2 ] \nonumber \\ 
& \qquad + \lambda_{\varphi S} |\varphi|^2 |S|^2 + \lambda_{ \varphi \Delta} |\varphi|^2 {\rm Tr}[\Delta^\dagger \Delta ] + \lambda_{S \Delta} |S|^2 {\rm Tr}[\Delta^\dagger \Delta]  
 + \left[\frac{\lambda_0}{2} H^T\epsilon \Delta^\dag H S + \rm{h.c.} \right],
\end{align}
where $\epsilon$ is two by two anti-symmetric matrix, the index $\alpha,i,j=1-3$ represents the number of family, $M$ can be diagonal without loss of generality, we omitted the terms containing the SM Higgs field in the potential, and all the Yukawa couplings to induce the SM fermion masses are the same as the SM.
The term $\lambda_0$ contribute to the one-loop induced neutrino masses.
Here we assume all the parameters above are positive real for simplicity.\\

\subsection{Higgs sector}
Here we formulate the Higgs sector.
First of all, we define each of the boson as follows:
 \begin{align}
 &H =\left[
\begin{array}{c}
w^+\\
\frac{v+h+iz}{\sqrt2}
\end{array}\right],\
\Delta_1=\left[
\begin{array}{cc}
\frac{ \delta^+}{\sqrt2} &\delta^{++}\\
\delta_0 & -\frac{ \delta^+}{\sqrt2} 
\end{array}\right],\
\varphi \equiv \frac{v'+\rho +iz'}{\sqrt2},
\label{component}
 \end{align}
where $H$ is the SM Higgs field, $v\approx246$ GeV,  and we assume the mixing between $\rho, h$ to be negligibly tiny that is in agreement with the current experimental results of the SM Higgs search at the LHC, while there exists nonzero mixing between $S,\delta_0$ due to $\lambda_0$. 
Here we define as follows:
\begin{align}
&\delta_0=c_\alpha H_1 + s_\alpha H_2,
\quad S=-s_\alpha H_1 + c_\alpha H_2,
\end{align}
where $s(c)_\alpha$, which is the short-hand notation of $\sin(\cos)\alpha$, is proportional to $\lambda_0$.

\begin{figure}[t]
\begin{center}
\includegraphics[width=8cm]{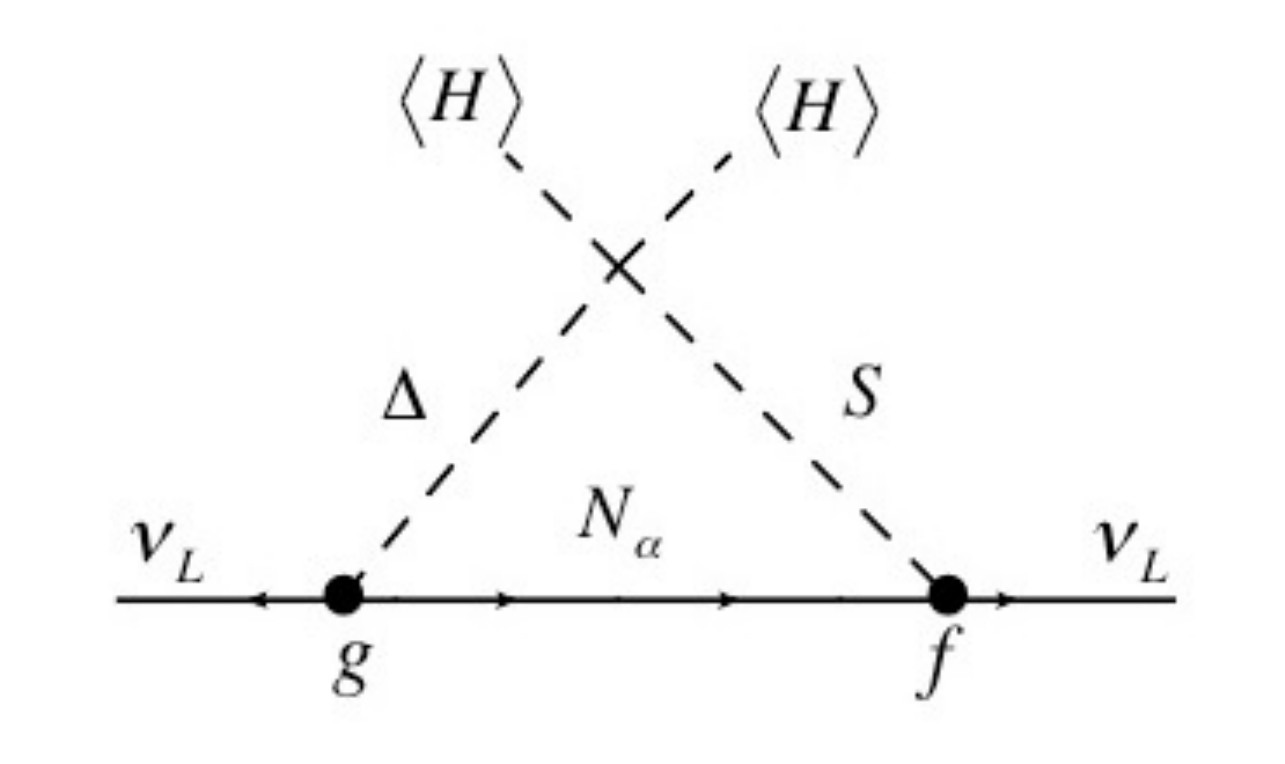}
\caption{The Feynman diagram for generating neutrino mass matrix.  } 
  \label{fig:neutrino}
\end{center}\end{figure}

\subsection{ Mass matrix for active neutrinos}
Here we formulate the active neutrino sector.
First of all, let us write the relevant Lagrangian in terms of mass eigenstate as follows:
\begin{align}
-{\cal L}_\nu&= f_{i\alpha}(\bar \ell_{L_i} E_{R_\alpha}+\bar \nu_{L_i} N_{R_\alpha}) S
+g_{\alpha j} \left(-\frac{1}{\sqrt2}\bar E^c_{L_\alpha} \nu^c_{L_j} \delta^- + \bar N_{L_\alpha} \nu^c_{L_j} \delta^*\right) \nonumber \\
&
-g_{\alpha j} \left(\frac{1}{\sqrt2}\bar N_{L_\alpha} \ell^c_{L_j} \delta^- + \bar E_{L_\alpha} \ell^c_{L_j} \delta^{--}\right)+{\rm h.c.},
\label{eq:Yukawa}
\end{align}
where interactions inducing the neutrino mass corresponds to the first line, while the second term contributes to the muon anomalous magnetic dipole moment as well as the lepton flavor violations (LFVs).
Then the  active neutrino mass matrix $m_\nu$
is given at one-loop level via two inert bosons as Fig.~\ref{fig:neutrino}, and its formula is given by
\begin{align}
&(m_{\nu})_{ij}
=-\frac{s_\alpha c_\alpha}{(4\pi)^2}
\sum_{\alpha}[f_{i\alpha}g_{\alpha j}+(f_{i\alpha}g_{\alpha j})^T] F_I(M_\alpha,m_{H_1},m_{H_2})\equiv 
-\sum_{\alpha}[f_{i\alpha} R_\alpha g_{\alpha j}+(f_{i\alpha} R_\alpha g_{\alpha j})^T],\\
& F_I(m_1,m_{2},m_{3}) \equiv
 \frac{m_1^2m_2^2 \ln\frac{m_1^2}{m_2^2}+m_1^2m_3^2 \ln\frac{m_3^2}{m_1^2}+ m_2^2m_3^2 \ln\frac{m_2^2}{m_3^2}}
 {(m_1^2-m_2^2)(m_1^2-m_3^2)},\
 R\equiv \frac{s_\alpha c_\alpha}{(4\pi)^2}  F_I(M_\alpha,m_{H_1},m_{H_2}),
\end{align}
Once we define $m_\nu\equiv V_{MNS} D_\nu V_{MNS}^T$, one can rewrite Yukawa coupling $f(g)$ in terms of several known parameters as follows~\cite{Okada:2015vwh}:
\begin{align}
f&=-\frac12[V_{MNS} D_\nu V_{MNS}^T+A] g^{-1} R^{-1}\ {\rm or}\
g=-\frac12 R^{-1} f^{-1}[V_{MNS} D_\nu V_{MNS}^T+A],
\end{align}
where $A$ is an arbitrary three by three anti-symmetric matrix satisfying $A^T+A=0$ with three complex values,
$V_{MNS},D_\nu$ is respectively lepton mixing and active neutrino masses that are observables~\cite{pdg}.
In our numerical convention, we choose the latter parametrization.

\subsection{ Lepton flavor violations (LFVs)}
LFV processes $\ell_i \to \ell_j \gamma$ arises from the Yukawa couplings $f$ and $g$,
and its formula of branching ratio can be given by~\cite{Lindner:2016bgg}
\begin{align}
& BR(\ell_i \to \ell_j \gamma)\approx \frac{48\pi^3 C_{ij}\alpha_{em}}{G_F^2 m^2_{\ell_i}}(|a_{R_{ij}}|^2+|a_{L_{ij}}|^2),\\
& a_{R_{ij}}\approx -\frac{m_{\ell_i}}{(4\pi)^2}\sum_{\alpha}
\left(
f_{j\alpha} f^\dag_{\alpha i} [s^2_\alpha G_1(m_{H_1},M_\alpha)+c^2_\alpha G_1(m_{H_2},M_\alpha)]\right.\nn\\
&\left.
-g^*_{j\alpha} g^T _{\alpha i} [2 G_1(M_\alpha,m_{\delta^{\pm\pm}}) +G_1(m_{\delta^{\pm\pm}},M_\alpha)+\frac12
(M_\alpha,m_{\delta^{\pm}})  ]
\right),\\
& a_{L_{ij}}\approx -\frac{m_{\ell_j}}{(4\pi)^2}\sum_{\alpha}
\left(
f_{j\alpha} f^\dag_{\alpha i} [s^2_\alpha G_1(m_{H_1},M_\alpha)+c^2_\alpha G_1(m_{H_2},M_\alpha)]\right.\nn\\
&\left.
-g^*_{j\alpha} g^T _{\alpha i} [2 G_1(M_\alpha,m_{\delta^{\pm\pm}}) +G_1(m_{\delta^{\pm\pm}},M_\alpha)+\frac12
(M_\alpha,m_{\delta^{\pm}})  ]
\right),\\
G_1(m_1,m_2)&=
\frac{2 m_1^6+3m_1^4m_2^2-6m_1^2m_2^4+m_2^6+12m_1^4m_2^2\ln\frac{m_2}{m_1}}{12(m_1^2-m_2^2)^4},
\end{align}
where $\ell_{e,\mu,\tau}\equiv (e,\mu,\tau)$, $\alpha_{em}\approx 1/137$, $G_F\approx1.17\times 10^{-5}$ GeV$^{-2}$, $C_{21}\approx1$,
$C_{31}\approx0.1784$, $C_{32}\approx0.1736$, and current experimental upper bounds are given by~\cite{pdg, TheMEG:2016wtm}:
\begin{align}
BR(\mu\to e\gamma)\lesssim 4.2\times 10^{-13},\quad BR(\tau\to e\gamma)\lesssim 3.3\times 10^{-8},\quad
BR(\tau\to \mu\gamma)\lesssim 4.4\times 10^{-8}.
\end{align}

\subsection{Muon anomalous magnetic dipole moment ($\Delta a_\mu$)} 
$\Delta a_\mu$ can easily be found from the formula of LFVs,
and its form is given by
\begin{align}
\Delta a_\mu\approx -m_\mu(a_{R_{\mu\mu}}+a_{L_{\mu\mu}}).
\end{align}
The experimental result for example  suggests the following bound~\cite{Hagiwara:2011af}:
\begin{align}
\Delta a_\mu=(26.1\pm8.0)\times 10^{-10}, \label{eq:dau-exp}
\end{align}
which is 3.3 $\sigma$ deviation from the SM contribution.
Clearly we expect $f\gg g$ in order to get positive $\Delta a_\mu$ in light of the experimental result.

\subsection{$Z'$ boson from $U(1)_H$ breaking}

We obtain massive $Z'$ boson after spontaneous $U(1)_H$ symmetry breaking by nonzero VEV of $\varphi$.
The mass of $Z'$ is given by
\begin{equation}
m_{Z'} = 4 x g' v' 
\end{equation}
where $g'$ is the $U(1)_H$ gauge coupling constant.
Our $Z'$ boson can decay into $L'_a$ as well as scalar bosons from $S$ and $\Delta$ as these particle have $U(1)_H$ charge.
Considering loop level, $Z'$ can also decay into the SM leptons and the decay width is given by~\cite{Ko:2017yrd}
\begin{align}
\Gamma(Z'\to\ell_i^+ \ell^-_j) \approx
\frac{m_{Z'}g'^2x^2}{8\pi(4\pi)^2}
\left|\sum_{\alpha} f_{i\alpha}f^\dag_{\alpha j}
\int[dx]_3\ln\left[\frac{x m_{H_2}^2 +(y+z) M^2_\alpha-yz m^2_{Z'}}{x M^2_\alpha +(y+z) m_{H_2}^2-yz m^2_{Z'}}\right]
\right|^2,
\label{eq:BR}
\end{align}
where $\ell_{i,j}$ involves charged-leptons and neutrinos; $(e,\mu,\tau,\nu_e,\nu_\mu,\nu_\tau)$ assuming to be their massless final state.
Here we have ignored the contribution from loop diagram associated with $\Delta$ since we assume relation $g_{ij} \ll f_{ij}$ to obtain positive $\Delta a_\mu$.
Notice that  $Z'$ dominantly decays into the SM lepton modes if $L'$, $S$ and $\Delta$ are sufficiently heavier than $m_{Z'}/2$.
Also $\Gamma(Z' \to \nu_i \bar \nu_j)$ is given by the same formula as Eq.~(\ref{eq:BR}). Therefore the branching ratios are given by
\begin{equation}
BR(Z' \to \ell^+_i \ell^-_j (\nu_i \bar \nu_j )) = \frac{1}{(1+ \delta_{ij})} \frac{ \Gamma(Z' \to \ell^+_i \ell^-_j) }{\sum_{i,j = e, \mu, \tau} \Gamma(Z' \to \ell^+_i \ell^-_j)},
\end{equation}
and $BR(Z' \to e^+ e^-)$ is maximally $1/2$.

\begin{figure}[t]
\begin{center}
\includegraphics[width=10cm]{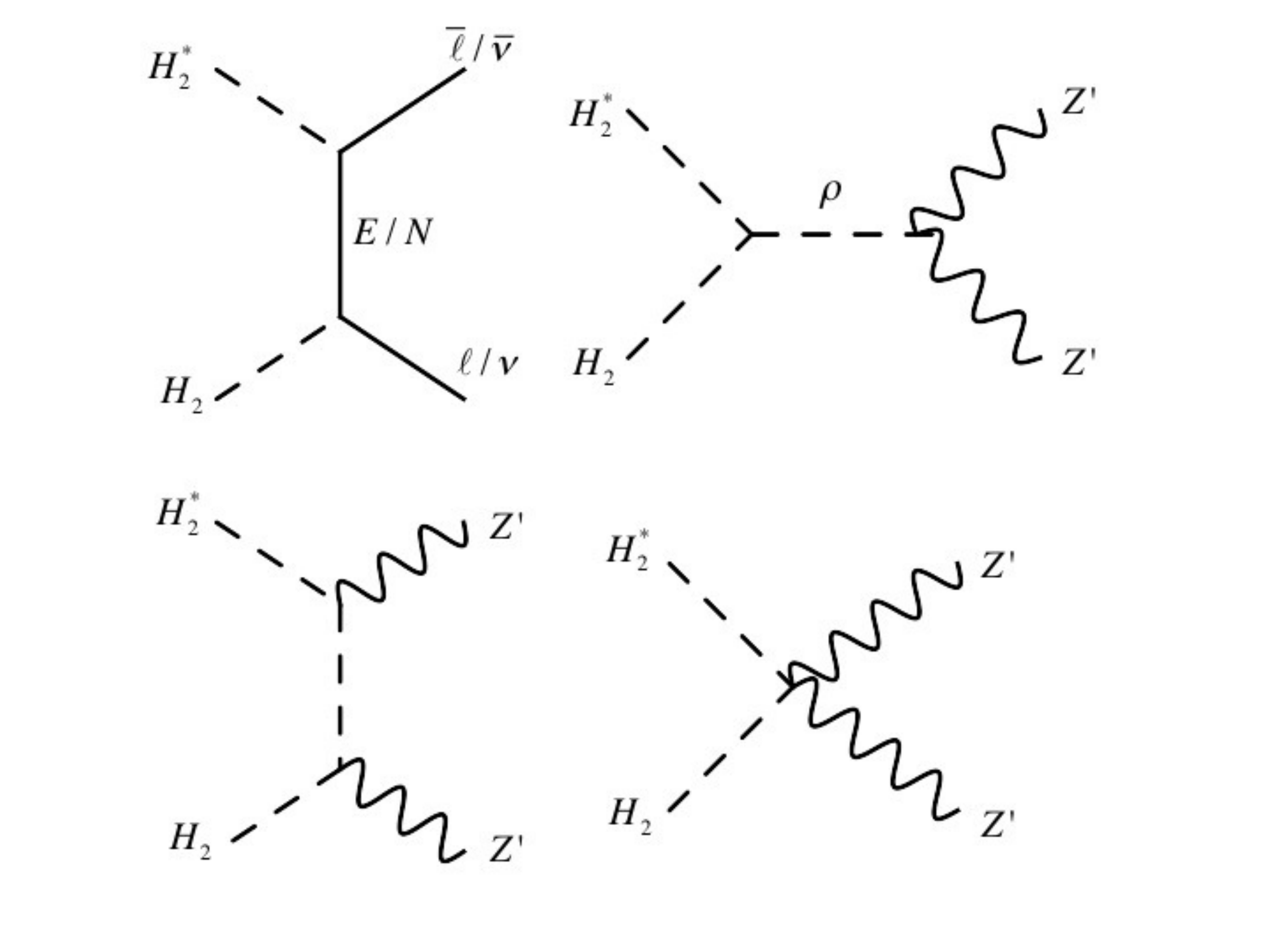}
\caption{The Feynman diagrams for DM annihilation processes.  } 
  \label{fig:DM}
\end{center}\end{figure}
\subsection{ The dark matter candidate and its interactions}
Firstly we identify the DM candidate as the $S$ dominated boson $H_2 \approx S$ as discussed above.
The relevant interactions for DM physics are given by
\begin{equation}
\mathcal{L} \supset -i x g' Z'^\mu (H_2^* \partial_\mu H_2 - H_2 \partial_\mu H_2^*) +  \frac{\lambda_{S\varphi} m_{Z'}}{4x g'} \rho H_2^*H_2  + g'^2 x^2 H_2^* H_2 Z'^\mu Z'_\mu+ 4 g' x m_{Z'} \rho Z'^\mu Z'_\mu,
\end{equation}
where we also have Yukawa interaction among DM and leptons as shown in Eq.~(\ref{eq:Yukawa}).
The possible DM annihilation processes are shown in Fig.~\ref{fig:DM}.
Among them the right-bottom diagram provides dominant contribution in our parameter setting; the left diagrams are suppressed since they are p- or d-wave contribution and we take $H_2 H_2^* \rho$ coupling is small so that we can evade the bound on direct detections. Therefore it should be of the order 0.01.
In our analysis below, we will numerically estimate the relic density of our DM using {micrOMEGAs 4.3.5}~\cite{Belanger:2014vza} by implementing relevant interactions.
The resulting cross section $H_2 H_2^* \to  \bar\ell_i\ell_j\bar\ell_{i'}\ell_{j'}$ is given by
\begin{align}
&\sigma v_{rel}(H_2H_2^*\to \bar\ell_i\ell_j\bar\ell_{i'}\ell_{j'} ) \approx\sigma v_{rel}(H_2H_2^*\to 2Z') \times 
BR(Z'\to \bar\ell_i\ell_j)BR(Z'\to \bar\ell_{i'}\ell_{j'}).
\end{align}
In our analysis, we estimate $BR(Z' \to \ell \ell')$ combining with the neutrino oscillation data and LFV constraints.

\begin{figure}[t]
\includegraphics[width=80mm]{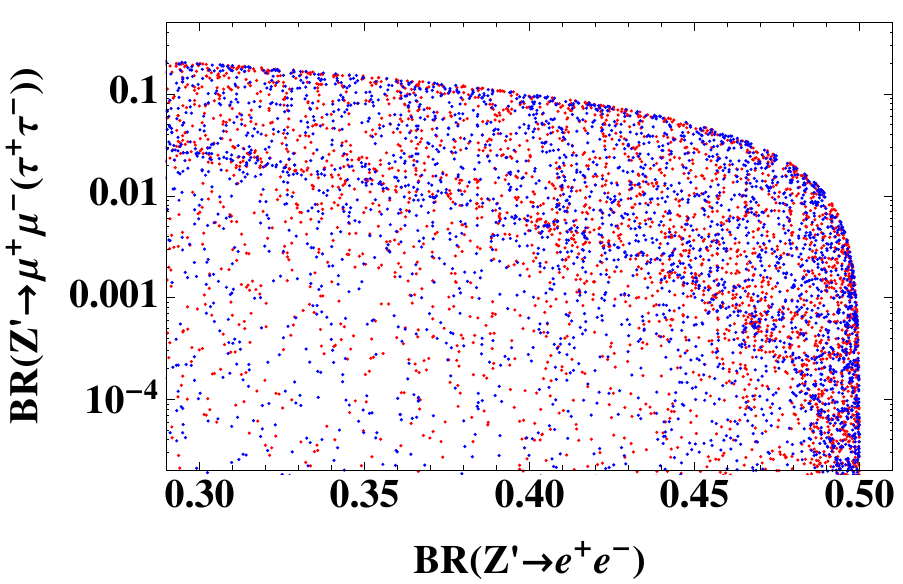}
\includegraphics[width=80mm]{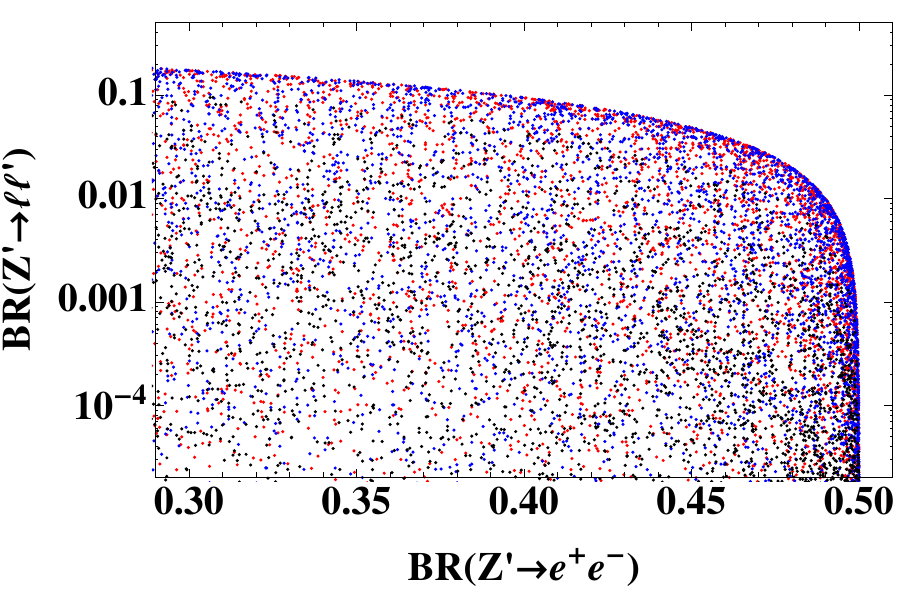}
\caption{
Left plot: Scatter plot of $BR(Z' \to e^+ e^-)$ versus $BR(Z' \to \mu^+ \mu^-)$ with blue points and $BR(Z' \to \tau^+ \tau^-)$ with red points
that satisfies neutrino oscillation data and LFV constraints. Right plot: Scatter plot of $BR(Z' \to e^+ e^-)$ versus $BR(Z' \to e \mu)$ with blue points, $BR(Z' \to e \tau)$ with red points
and $BR(Z' \to \mu \tau)$ with black points.}
\label{fig:BRs}
\end{figure}
\begin{figure}[t]
\includegraphics[width=80mm]{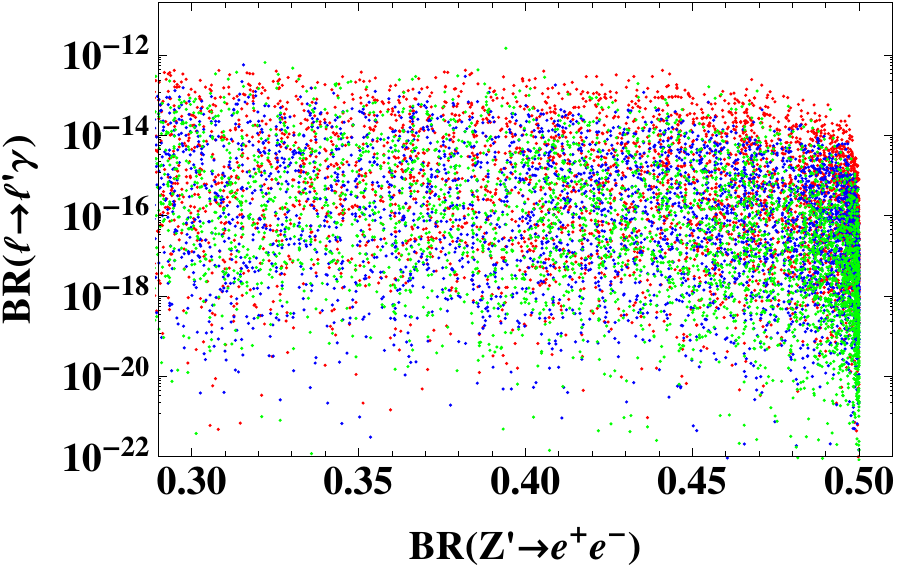}
\includegraphics[width=80mm]{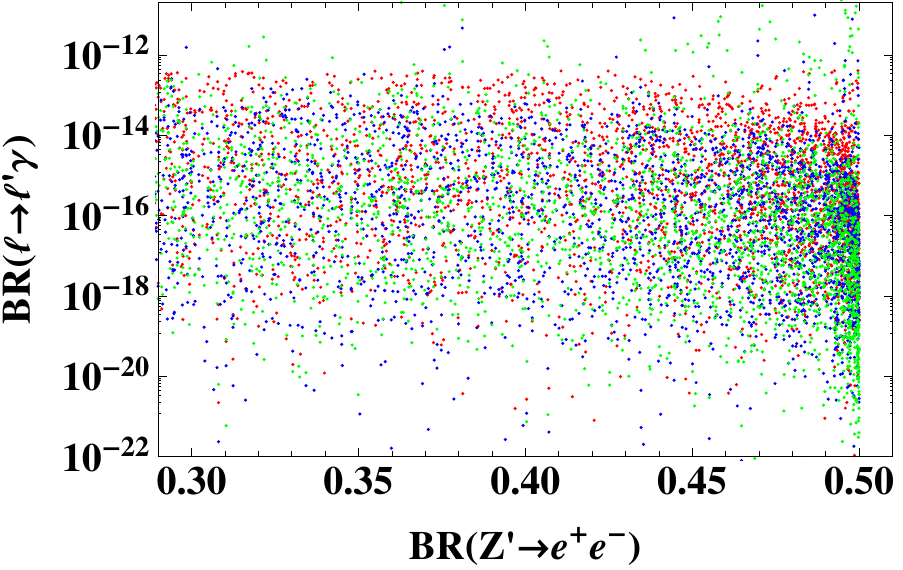}
\caption{
Scatter plots on $BR(Z' \to e^+ e^-)$-$BR(\ell \to \ell' \gamma)$ plane for NO(IO) cases in left(right) plot where red, blue and green points correspond to $BR(\mu \to e \gamma)$, $BR(\tau \to e \gamma)$ and  $BR(\tau \to \mu \gamma)$.}
\label{fig:BRLFV}
\end{figure}

\section{Numerical analysis \label{sec:numerical}}
In this section, we carry out numerical calculations and investigate possible explanation of DAMPE data. 
First of all we fix the following parameters to be $m_{H_2}=3000$ GeV to fit the DAMPE data to explain the excess around $\sim 1.4$ TeV,  $s_\alpha=0.3$, $m_{\delta^{\pm\pm}}=m_{\delta^{\pm}}=m_\delta$~\footnote{These mass degenerates provide the satisfying region of the oblique parameters.} in the numerical analysis. Then we select $1.5 \times10^4$ random sampling points for the valid input 
parameters in the following ranges:
\begin{align}
&(|A_{12}|,|A_{13}|,|A_{23}|) \in  {[10^{-15},10^{-10}]\ \text{GeV}},\
(M_{1} \le M_{2}\le M_{3}) \in {[ m_{H_2},5000]\ \text{GeV}},\nn\\ 
&(m_{H_1},m_\delta) \in  {[ m_{H_2},5000]\ \text{GeV}}, \quad  |f_{ij}| \in   [10^{-3},1],\nn
\label{range_scanning}
\end{align}
and we search for allowed points in our parameter space that satisfy neutrino oscillation data and LFV constraints, where $g<\sqrt{4\pi}$ is imposed to satisfy the perturbation limit. 
Note that magnitude of Yukawa couplings $|g_{ij}|$ are found to be much smaller than $|f_{ij}|$ in the setup which is consistent with our assumption in above discussions.
For neutrino oscillation data, we apply best fit values for normal ordering (NO) and inverted ordering (IO) cases~\cite{pdg}.
 Here we take range of Yukawa couplings $f_{ij}$ universally to see behavior of branching ratio of $BR(Z'\to \ell^\pm \ell'^\mp)$ taking into account the constraints from neutrino oscillation data and LFV constraints.
%
In the left panel of Fig.~\ref{fig:BRs}, we show scatter plots of $BR(Z' \to e^+ e^-)$ versus $BR(Z' \to \mu^+ \mu^-)$ with blue points and $BR(Z' \to \tau^+ \tau^-)$ with red points
that satisfies neutrino oscillation data and LFV constraints. Here we focus on the region $BR(Z' \to e^+ e^-) \gtrsim 0.3$ since it is favored to explain the DAMPE excess.
We see slight correlation between the branching ratios but the $BR(Z' \to \mu^+ \mu^-(\tau^+ \tau^-))$ can take wide range of value. Also in the right panel of Fig.~\ref{fig:BRs},  we show 
scatter plot of $BR(Z' \to e^+ e^-)$ versus $BR(Z' \to e \mu)$ with blue points, $BR(Z' \to e \tau)$ with red points and $BR(Z' \to \mu \tau)$ with black points. The figure shows $BR(Z' \to \tau \mu)$ tends to smaller than the others for relatively large $BR(Z' \to e^+ e^-)$ region. In addition, we find the results are same for NO and IO cases.

In Fig.~\ref{fig:BRLFV}, we show the 
scatter plot on $\{ BR(Z' \to e^+ e^-), BR(\ell \to \ell' \gamma) \}$ plane where $BR(\mu \to e \gamma)$, $BR(\tau \to e \gamma)$ and $BR(\tau\to \mu\gamma)$ correspond to red, blue and green plots, parameters for these points satisfy neutrino oscillation data and LFV constraints, and left-(right-)plot is for NO(IO) case.
We find that the $BR(\ell\to \ell'\gamma)$ in case of NO tends to be smaller than the one in case of IO, and
$BR(\mu\to e\gamma)$ reach the current experimental upper bound. Therefore $BR(\mu\to e\gamma)$ could be tested near future~\cite{TheMEG:2016wtm}.
As for another issue, the scale of $\Delta a_\mu$ is at most $10^{-12}$ , which is $10^3$ times as same as the measured value, in obtaining sizable $BR(Z' \to e^+ e^-)$.
This tendency is due to the fact that we need to suppress Yukawa couplings $f_{ij}$ associated with muon index compared to those with electron index. Thus if we do not require sizable $BR(Z' \to e^+ e^-)$, it is possible to enhance $\Delta a_\mu$.
\footnote{If we make the Yukawa couplings $|f_{21}|,|f_{22}|,|f_{23}|$ to be large as $ f_{2i} \gtrsim \mathcal{O}(1)$, one finds the sizable $\Delta a_\mu$.}

\begin{figure}[t]
\includegraphics[width=110mm]{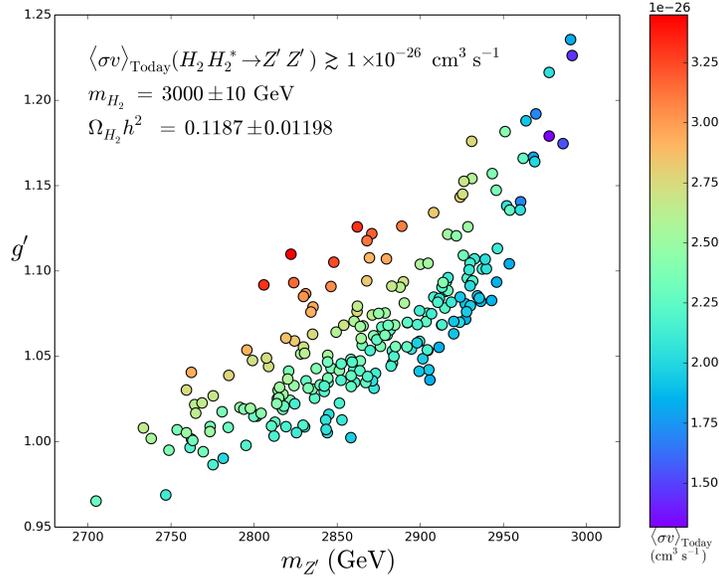}
\caption{
Scatter plot of $m_{Z'}$ versus $g'$ with color indicating $\sigma v_{rel}(H_2H_2^*\to 2Z')\gtrsim (1.25,\, 3.5)\times 10^{-26} \,{\rm cm^3/s}$ in current universe. See the main text for more details.}
\label{fig:DAMPE-mZp-gp}
\end{figure}

In Fig.~\ref{fig:DAMPE-mZp-gp}, we show the scatter plot of $m_{Z'}$ versus $g'$ with color indicating $\sigma v_{rel}(H_2H_2^*\to 2Z')= (1.25,\, 3.5)\times 10^{-26} \,{\rm cm^3/s}$ in current universe. As a conservative estimation, $BR(Z' \to e^+ e^-) \gtrsim 0.3$ highlighted in the above analysis corresponds to $\sigma v_{rel}(H_2H_2^*\to e^+e^- \bar\ell_{i'}\ell_{j'} ) \gtrsim \sigma v_{rel}(H_2H_2^*\to 2Z') \Big(1-(1-0.3)^2\Big) \simeq (0.6,\, 1.7)\times 10^{-26} \,{\rm cm^3/s}$, half of which cover the region which can interpret the DAMPE peak, i.e. at least one $e^+e^-$ pair produced with $\gtrsim 1 \times 10^{-26} \,{\rm cm^3/s}$. When including other $e^+e^-$ produced from $Z' \to \mu^+\mu^-,\tau^+\tau^-$ decay modes, the $e^+e^-$ flux can be further increased, although with the spectrum slightly flattened.
Note also that we can have resonant enhancement of annihilation cross section if we include process $H_2 H_2^* \to \rho \to Z' Z'$ with sizable $\lambda_{S \varphi}$ coupling and tuning the mass relation as $2 m_{H_2} \sim m_\rho$.

\section{Conclusions}
We have studied a model with hidden gauge symmetry $U(1)_H$ where a bosonic DM candidate $H_2$ plays an role in inducing the active neutrino masses, lepton flavor violations, and muon anomalous magnetic moment at one-loop level.
In addition, $Z'$ boson from $U(1)_H$ decays into the SM leptons through one-loop effect where our DM candidate propagates inside a loop diagram.
Then we have shown branching ratio of $Z'$ in parameter region which can fit the neutrino oscillation data and satisfy constraints from LFVs. 
Moreover correlation between branching ratios of $Z'$ and LFV charged lepton decays has been investigated and different behavior has been shown in normal and inverted ordering cases for neutrino masses.

The relic density of DM has been explained from a contact interaction via $Z'$ boson inducing annihilation process of $H_2 H_2^* \to Z'Z'$. 
Then $Z'$ boson decays into leptons, giving four-body lepton final states; $H_2 H_2^*\to 2Z'\to \bar\ell\ell\bar\ell\ell$ ($\ell=e,\mu,\tau,\nu_e,\nu_\mu,\nu_\tau$).
We thus have found that the DAMPE excess can be accommodated when $Z'$ dominantly decays into electron positron pair, and masses of DM and $Z'$ is around 3 TeV.
In such a case, the scale of $\Delta a_\mu$ is at most $10^{-12}$, which is $10^3$ times as same as the measured value due to the electron specific Yukawa couplings. 
We note however that sizable $\Delta a_\mu$ can be obtained by changing the several scales of components of $f_{ij}$ associated with muon.

\section*{Acknowledgments}
\vspace{0.5cm}
We would like to thank Junjie Cao, Liangliang Shang and Xiaofei Guo for helpful discussions.

\end{document}